\documentclass[prb,twocolumn,superscriptaddress,showpacs]{revtex4}
\usepackage{graphicx}
\usepackage{dcolumn}
\usepackage{bm}

\begin{document}

\preprint{APS/123-QED}

\title{High energy spin excitations in BaFe$_{2}$As$_{2}$}

\author{R. A. Ewings}
\affiliation{ISIS Facility, Rutherford Appleton Laboratory, Chilton,
Didcot, OX11 0QX, UK}
\author{T. G. Perring}
\affiliation{ISIS Facility, Rutherford Appleton Laboratory, Chilton,
Didcot, OX11 0QX, UK}
\author{R. I. Bewley}
\affiliation{ISIS Facility, Rutherford Appleton Laboratory, Chilton,
Didcot, OX11 0QX, UK}
\author{T. Guidi}
\affiliation{ISIS Facility, Rutherford Appleton Laboratory, Chilton,
Didcot, OX11 0QX, UK}

\author{M. J. Pitcher}
\affiliation{ Department of Chemistry, Oxford University, Inorganic
Chemistry Laboratory, Oxford, OX1 3QR, UK}
\author{D. R. Parker}
\affiliation{ Department of Chemistry, Oxford University, Inorganic
Chemistry Laboratory, Oxford, OX1 3QR, UK}
\author{S. J. Clarke}
\affiliation{ Department of Chemistry, Oxford University, Inorganic
Chemistry Laboratory, Oxford, OX1 3QR, UK}
\author{A. T. Boothroyd}
\email{a.boothroyd@physics.ox.ac.uk} \affiliation{ Department of
Physics, Oxford University, Clarendon Laboratory, Oxford, OX1 3PU,
UK}

\date{\today}

%


\begin{abstract}
We report neutron scattering measurements of cooperative spin
excitations in antiferromagnetically ordered BaFe$_{2}$As$_{2}$, the
parent phase of an iron pnictide superconductor. The data extend up
to $\sim$ $100$\,meV and show that the spin excitation spectrum is
sharp and highly dispersive. By fitting the spectrum to a linear
spin-wave model we estimate the magnon bandwidth to be in the region
of 0.17\,eV. The large characteristic spin fluctuation energy
suggests that magnetism could play a role in the formation of the
superconducting state.
\end{abstract}


\maketitle

One of the greatest challenges presented by the recently discovered
iron pnictide superconductors \cite{Kamihara-JACS-2008} is to
identify the electron pairing interaction which permits the
formation of a superconducting condensate.  In conventional
superconductors this interaction is provided by the exchange of a
phonon. For the iron pnictides, however, theoretical calculations
\cite{Boeri-PRL-2008, Haule-PRL-2008} indicate that the
electron--phonon coupling is too weak to account for the observed
high critical temperatures. Attention has therefore turned to other
types of bosonic excitations which could mediate the pairing
interaction.

One such candidate is spin fluctuations \cite{Singh-PRL-2008}. In
common with the layered cuprates, superconductivity in the pnictides
is found in close proximity to parent phases which exhibit
long-range antiferromagnetic order
\cite{Dong-EPL-2008,Cruz-Nature-2008}. However, unlike the cuprates,
whose magnetic properties are governed by strong superexchange
interactions between localized spin--$\frac{1}{2}$ moments in a
single Cu 3$d_{x^2-y^2}$ orbital, magnetism in the pnictides is more
itinerant in character and derives from multiple $d$ orbitals. It
may also involve a degree of frustration. In magnetically ordered
materials the dominant magnetic excitations are coherent spin waves.
Wavevector-resolved measurements of the spin-wave spectrum by
inelastic neutron scattering provide information on the fundamental
magnetic interactions and can also reveal effects due to itinerancy
and frustration. Such studies on the magnetically ordered parent
phases of unconventional superconductors like the cuprates and iron
pnictides are important to establish the characteristic energy
scales of the spin fluctuations and also to provide a reference
against which changes associated with superconductivity can be
identified.

Here we present neutron scattering data on the collective spin
excitations in antiferromagnetic BaFe$_{2}$As$_{2}$. We find that
the spin excitation spectrum has a very steep dispersion within the
FeAs layers with a bandwidth in the region of 0.17\,eV, not much
less than that in the cuprates. Such a high characteristic energy
suggests that spin fluctuations are a serious candidate to mediate
high temperature superconductivity in the iron pnictides.

\begin{figure}
\begin{center}
\includegraphics
[width=8cm,bbllx=20,bblly=20,bburx=574, bbury=390,angle=0,clip=]
{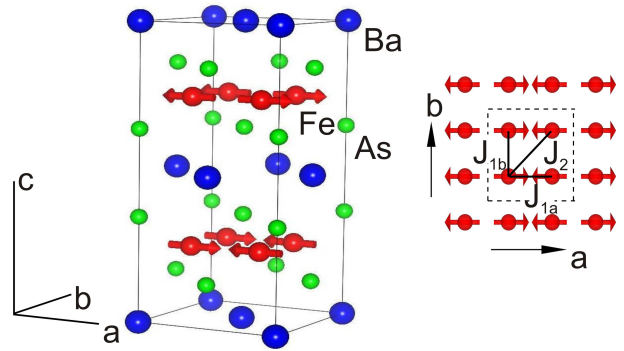} \caption{\label{fig1} (Color online) Crystal
and magnetic structure of BaFe$_{2}$As$_{2}$. On the left is the
crystal structure with the conventional unit cell for the low
temperature orthorhombic $Fmmm$ structure. The three--dimensional
antiferromagnetic (AFM) ordering of Fe spins is indicated. On the
right is a single layer of Fe spins showing the in-plane AFM order
and the nearest- and next-nearest-neighbour exchange interactions. }
\end{center}
\end{figure}

The parent phase BaFe$_{2}$As$_{2}$ becomes superconducting on
doping with holes \cite{Rotter-arXiv-2008} or on application of
pressure \cite{Alireza-arXiv-2008}. At $T_{\rm s} = 140$\,K,
BaFe$_{2}$As$_{2}$ undergoes a structural transition from tetragonal
to orthorhombic and simultaneously develops three-dimensional
long-range antiferromagnetic order \cite{Rotter-PRB-2008,
Huang-arXiv-2008, Su-arXiv-2008}. On cooling through $T_{\rm s}$,
the space group changes from $I4/mmm$ (lattice parameters $a=
3.96$\,{\AA}, $c = 13.0$\,{\AA}) to $Fmmm$ (lattice parameters $a=
5.61$\,{\AA}, $b= 5.57$\,{\AA}, $c = 12.9$\,{\AA}). The magnetic
structure of BaFe$_{2}$As$_{2}$, shown in Fig.\ \ref{fig1}, is a
collinear antiferromagnet with propagation vector ${\bf Q}_{\rm AF}
= (1, 0, 1)$. The ordered moments on the Fe atoms are of approximate
magnitude 0.9\,$\mu_{\rm B}$ and point along the orthorhombic $a$
axis.

Polycrystalline BaFe$_{2}$As$_{2}$ was prepared by reacting
stoichiometric amounts of the elements in a tantalum ampoule sealed
under argon (800$^\circ$C for 2 days, then 900$^\circ$C for 2 days
after regrinding). Phase purity was confirmed using X-ray powder
diffraction. The neutron scattering experiments were performed on
the MERLIN chopper spectrometer at the ISIS Facility
\cite{Bewley-PhysicaB-2006}. Approximately 8\,g of the
BaFe$_2$As$_2$ powder was sealed inside a cylindrical aluminium can
mounted in a top-loading closed-cycle refrigerator. Spectra were
recorded at a temperature of 7\,K with four different neutron
incident energies: $E_{\rm i} = 25$\,meV, 50\,meV, 200\,meV and
400\,meV. The scattering from a standard vanadium sample was used to
normalize the spectra and to place them on an absolute intensity
scale with units mb\,sr$^{-1}$\,meV$^{-1}$\,f.u.$^{-1}$, where f.u.\
stands for `formula unit' (of BaFe$_{2}$As$_{2}$). The spectra were
azimuthally-averaged and transformed onto a $(Q,{\rm energy})$ grid,
where $Q = |{\bf Q}|$ is the magnitude of the neutron scattering
vector. The presented intensity is the partial differential
cross-section ${\rm d}^2\sigma /{\rm d}\Omega {\rm d}E_{\rm f}$
multiplied by the factor $k_{\rm i}/k_{\rm f}$ \cite{Squires}, where
$k_{\rm i}$ and $k_{\rm f}$ are the initial and final neutron
wavevectors and $E_{\rm f}$ is the final neutron energy.

\begin{figure}
\includegraphics
[width=7.5cm,bbllx=20,bblly=20,bburx=534, bbury=823,angle=0,clip=]
{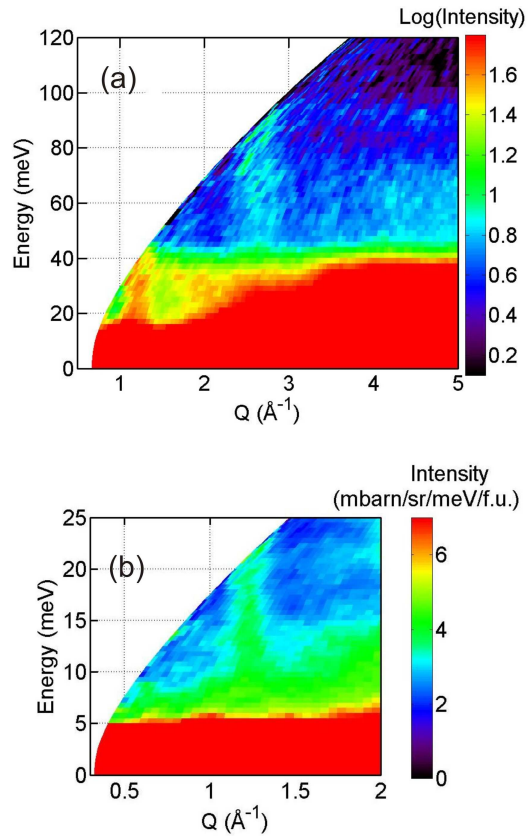} \caption{\label{fig2} (Color online) Neutron
scattering spectra of BaFe$_{2}$As$_{2}$. The data were recorded at
a temperature of 7\,K with incident neutron energies of (a) 200\,meV
and (b) 50\,meV. The pillars of scattering centred near
$Q=1.2$\,{\AA}$^{-1}$ and $Q=2.6$\,{\AA}$^{-1}$ are caused by highly
dispersive collective excitations associated with the
antiferromagnetically ordered Fe spins (see Fig.\ \ref{fig1}). (b)
is a higher resolution image showing the magnetic signal emerging
from $Q=1.2$\,{\AA}$^{-1}$.  }
\end{figure}

Figure \ref{fig2}(a) illustrates the general features of the data.
At low energies there is strong diffuse scattering due to the
elastic peak and scattering from phonons, the latter of which
increases with $Q$. The phonon signal drops off sharply above
40\,meV, which is the upper limit of the vibrational density of
states \cite{Mittal-PRB-2008}. Two distinct features stand out from
the phonon signal. One is a narrow pillar of scattering at $Q
=1.2$\,{\AA}$^{-1}$, and the second is a plume of intensity centred
on $Q=2.6$\,{\AA}$^{-1}$. The latter extends in energy to at least
90\,meV where it disappears out of the accessible region of $(Q,{\rm
energy})$ space. The $1.2$\,{\AA}$^{-1}$ feature is followed to
lower energies in Fig.\ \ref{fig2}(b), which was obtained with a
higher resolution configuration.

The origin of the $1.2$\,{\AA}$^{-1}$ and $2.6$\,{\AA}$^{-1}$
features is the cooperative spin wave excitations (magnons)
associated with the antiferromagnetic (AFM) zone centres $(1,0,\,l)$
and $(1,2,\,l)$. Since the structure is layered we expect only a
weak variation in the inelastic scattering with $l$. The effect of
powder-averaging and resolution-folding makes the 2D magnon
scattering appear at slightly higher $Q$ than the 2D AFM wavevectors
$Q_{(1,0)} = 1.1$\,{\AA}$^{-1}$, $Q_{(1,2)} = 2.5$\,{\AA}$^{-1}$, as
observed in Fig.\ \ref{fig2}.

\begin{figure}
\includegraphics
[width=8.0cm,bbllx=236,bblly=346,bburx=594, bbury=742,angle=0,clip=]
{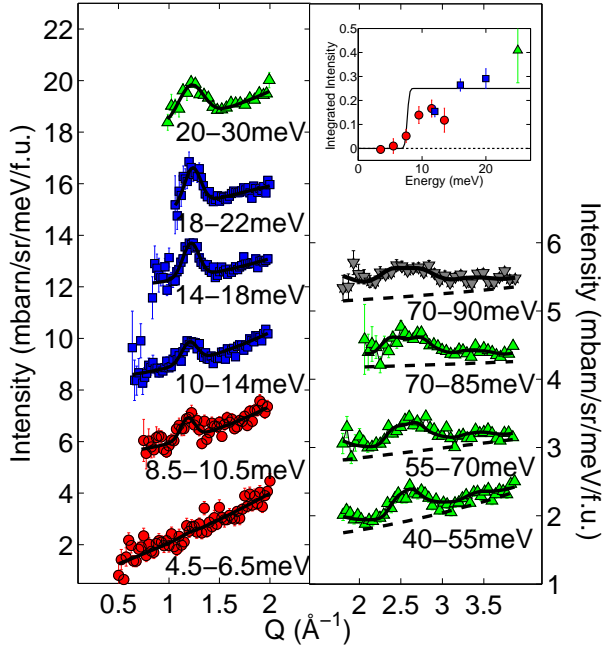} \caption{\label{fig3} (Color online) The
left and right panels show a series of constant-energy cuts through
the magnon signals at $Q=1.2$\,{\AA}$^{-1}$ and
$Q=2.6$\,{\AA}$^{-1}$, respectively. The data were averaged over the
energy ranges indicated. Successive cuts are displaced vertically
for clarity. The symbols represent different neutron incident
energies: 25\,meV (red circles), 50\,meV (blue squares), 200\,meV
(green triangles), 400\,meV (grey inverted triangles). In the left
panel the lines are fits to Gaussian peaks on a sloping background.
In the right panel the lines are constant-energy cuts (averaged over
the same energy ranges as the data) through the powder-averaged spin
wave spectrum calculated with parameters $SJ_2=2SJ_{1a}=2SJ_{1b} =
35$\,meV, $J_c=0$, $SK_{ab}=SK_c=0.042$\,meV and $S_{\rm eff}=0.28$
(see Fig.\ \ref{fig4}). The insert shows the integrated intensities
of the magnon scattering measured at $Q=1.2$\,{\AA}$^{-1}$ as a
function of energy. The line indicates the expected step in
intensity for a clean gap folded with the experimental resolution.}
\end{figure}

Figure\ \ref{fig3} shows examples of a series of cuts taken through
the data at different energies.  At the higher energies the signal
is seen to broaden (right panel). This is due to dispersion of the
spin waves. Below $\sim$15\,meV the magnetic signal decreases in
intensity. We fitted the cuts taken through $Q =1.2$\,{\AA}$^{-1}$
with a Gaussian line shape on a quadratic background and plot the
integrated intensities of the fitted peaks at each energy in the
insert to Fig.\ \ref{fig3}. The data show that the magnetic
excitations are gapped, with no detectable signal below 5\,meV.
However, the gap is not sharp since a sharp gap would produce the
resolution-broadened step in energy shown in Fig.\ \ref{fig3}. The
broadening of the step could be due to dispersion of the gap in the
$c$ direction and/or the existence of two or more gaps in the
5--15\,meV range. In orthorhombic symmetry two gaps are expected at
${\bf Q}_{\rm AF}$ as a result of magnetic anisotropy which splits
the spin waves into two non-degenerate branches with predominantly
in-plane and out-of-plane character, respectively.

We now compare our data with a linear spin wave model for
BaFe$_{2}$As$_{2}$ based on an effective Heisenberg spin
Hamiltonian. The suitability of such a model is perhaps questionable
in view of the itinerant character of the magnetism
\cite{Mazin-arXiv-2008}, but at the very least it will provide an
estimate of the scale of the magnetic interactions. We calculated
the spectrum by the same method as Yao and Carlson
\cite{Yao-arXiv-2008} but extended the Hamiltonian to include terms
that represent the single-ion anisotropy:
\begin{equation}
H = \sum_{\langle jk\rangle}J_{jk}{\bf S}_j \cdot {\bf S}_k +
\sum_{j}\{K_c(S_z^2)_j + K_{ab}(S_y^2-S_x^2)_j\}.\label{eq1}
\end{equation}
The first summation is over nearest-neighbour and
next-nearest-neighbour pairs with each pair counted only once. The
$J_{jk}$ are exchange parameters as defined in Fig.\ \ref{fig1}, and
$K_{ab}$ and $K_c$ are in-plane and out-of-plane anisotropy
constants, respectively. Diagonalisation of equation (\ref{eq1})
leads to two non-degenerate branches with dispersion
\begin{equation}
\hbar\omega_{1,2}({\bf Q}) = \sqrt{A_{\bf Q}^2-(C\pm D_{\bf
Q})^2},\label{eq2}
\end{equation}
where
\begin{eqnarray}
A_{\bf Q} & = & 2S\{J_{1b}[\cos(\frac{{\bf Q}\cdot{\bf b}}{2})-1]+
J_{1a} + 2J_2+J_c\} \nonumber\\[2pt]
& & \hspace{80pt} + S(3K_{ab}+K_c)\nonumber\\[5pt] C &
= & S(K_{ab}-K_c)\nonumber\\[5pt] D_{\bf Q} & = & 2S\{J_{1a}\cos(\frac{{\bf Q}\cdot{\bf
a}}{2})+2J_2\cos(\frac{{\bf Q}\cdot{\bf a}}{2})\cos(\frac{{\bf
Q}\cdot{\bf b}}{2}) \nonumber\\[2pt]
& & \hspace{80pt} + J_c\cos({\bf Q}\cdot{\bf c})\}.\label{eq3}
\end{eqnarray}

The neutron scattering cross section may be written \cite{Squires}
\begin{eqnarray}
\frac{{\rm d}^2\sigma}{{\rm d}\Omega{\rm d}E_{\rm f}} & = &
\frac{k_{\rm f}}{k_{\rm i}}\left(\frac{\gamma r_0}{2}
\right)^{\hspace{-2pt}2} g^2f^2(Q)\exp(-2W) \nonumber\\[5pt] & & \hspace{20pt}\times \sum_{\alpha
\beta}(\delta_{\alpha
\beta}-\hat{Q}_{\alpha}\hat{Q}_{\beta})S^{\alpha\beta}({\bf
Q},\omega),\label{eq4}
\end{eqnarray}
where $(\gamma r_0/2)^2 = 72.8$\,mb, $g$ is the $g$-factor of iron,
$f(Q)$ the form factor of iron, $\exp(-2W)$ is the Debye--Waller
factor which is close to unity at low temperatures,
$\hat{Q}_{\alpha}$ is the $\alpha$ component of a unit vector in the
direction of $\bf Q$, and $S^{\alpha\beta}({\bf Q},\omega)$ is the
response function describing $\alpha\beta$ spin correlations. Only
the transverse correlations ($yy$ and $zz$ for BaFe$_2$As$_2$)
contribute to the linear spin wave cross section and the response
functions (per BaFe$_{2}$As$_{2}$ formula unit) for magnon creation
are given by
\begin{eqnarray} S^{yy}({\bf Q},\omega) & =
& S_{\rm eff}\frac{A_{\bf Q}-C-D_{\bf Q}}{\hbar\omega_{1}({\bf Q})}
\{n(\omega)+1\}\delta[\omega-\omega_{1}({\bf Q})] \nonumber\\[5pt] S^{zz}({\bf Q},\omega) & =
& S_{\rm eff}\frac{A_{\bf Q}+C-D_{\bf Q}}{\hbar\omega_{2}({\bf Q})}
\{n(\omega)+1\}\delta[\omega-\omega_{2}({\bf Q})]\nonumber\\[5pt]
\end{eqnarray}
where $S_{\rm eff}$ is the effective spin and $n(\omega)$ is the
boson occupation number. In linear spin wave theory $S_{\rm eff} =
S$, but we keep them distinct here because in the analysis they are
obtained essentially independently (see below).


Because our data do not extend over the full spin wave dispersion we
were unable to determine $J_{1a}$, $J_{1b}$ and $J_2$ independently.
However, several authors
\cite{Si-PRL-2008,Yildirim-PRL-2008,Ma-arXiv-2008,Yin-PRL-2008} have
made predictions of effective Heisenberg exchange parameters from
first-principles electronic structure calculations. In some
cases\cite{Si-PRL-2008,Yildirim-PRL-2008,Ma-arXiv-2008} $J_2$ is
predicted to exceed $J_{1a}$ and $J_{1b}$ by about a factor of two,
with $J_{1a}$ and $J_{1b}$ either both ferromagnetic or both AFM
(providing $J_{1a}$ and $J_{1b}$ are the same sign the spectrum at
low energies is not very sensitive to which sign it is
\cite{Yao-arXiv-2008}). Alternatively\cite{Yin-PRL-2008},
$J_{1a} \gg J_{1b}$ and $J_{1a} \simeq 2J_2$. Guided by these
predictions we fixed the ratios of the exchange parameters to be
either (i) $J_{1a} = J_{1b} = J_2/2$, or (ii) $J_{1a} = 2J_2 =
-5J_{1b}$. Since we cannot resolve more than one gap in the data we
set $K_{ab}=K_c$ so that the in-plane and out-of-plane gaps are the
same, and we neglected the $c$-axis coupling, which is expected to
be much smaller than the in-plane coupling. The $g$-factor was set
to 2.

We computed the powder-averaged spin wave spectrum convoluted with
the instrumental resolution and fitted it to the experimental data
near $Q=2.6$\,{\AA}$^{-1}$ allowing only $SJ_2$ and $S_{\rm eff}$ to
vary. $S_{\rm eff}$ is essentially determined by the absolute
intensity, and $SJ_2$ by the dispersion. Data near
$Q=1.2$\,{\AA}$^{-1}$ were excluded as the peak widths are dominated
by instrument resolution and consequently are insensitive to $SJ_2$.
We obtained equally good fits to the data with both sets of
parameter ratios. The best-fit parameters are, for case (i) $SJ_2=
35 \pm 3$\,meV and $S_{\rm eff}=0.28 \pm 0.04$, and for case (ii)
$SJ_2= 18 \pm 1$\,meV and $S_{\rm eff}=0.54 \pm 0.05$. The highest
resolution data reveals a gap of $7.7 \pm 0.2$\, meV
--- Fig.\ \ref{fig3} (insert) --- which yields
$SK_c=SK_{ab}=0.042 \pm 0.004$\,meV [case (i)] and $0.053 \pm
0.005$\,meV [case (ii)]. For case (i) we tried different values of
$J_1/J_2$, but the best-fit value of $SJ_2$ was relatively
insensitive, varying from 33\,meV ($J_1/J_2=0.25$) to 46\,meV
($J_1/J_2=1$). Aside from the uncertainty in $J_1/J_2$, the major
error in $SJ_2$ is the statistical error on the fit. The error in
$S_{\rm eff}$ comes from estimates of the background and
instrumental resolution.

\begin{figure}
\includegraphics
[width=7.5cm,bbllx=32,bblly=167,bburx=544, bbury=585,angle=0,clip=]
{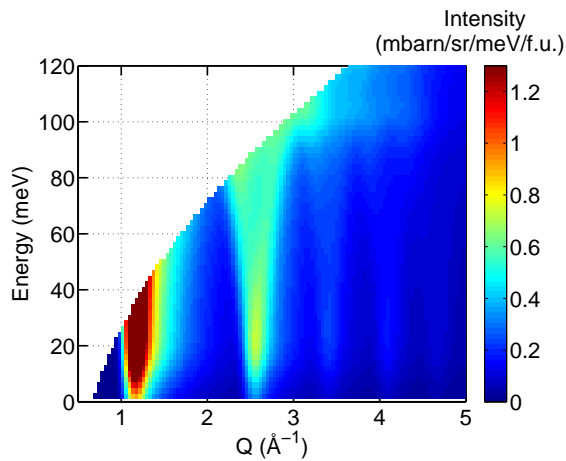} \caption{\label{fig4} (Color online)
Simulation of the powder-averaged spin wave spectrum of
BaFe$_{2}$As$_{2}$. The simulation covers the same $Q$ and energy
range as the data in Fig.\ \ref{fig2}(a). The parameters used for
the simulation are given in the caption to Fig.\ \ref{fig3}.}
\end{figure}

To illustrate the level of agreement, Fig.\ \ref{fig4} shows the
powder-averaged spin wave spectrum calculated with the best-fit
parameters for case (i) over the same range of $Q$ and energy
covered by the measurements in Fig.\ \ref{fig2}(a). The maximum spin
wave energy is $\sim$175 meV for both parameter sets, well out of
range of the present experiment. We note that the order of magnitude
of the experimentally-determined exchange parameters is consistent
with theoretical predictions
\cite{Yildirim-PRL-2008,Yin-PRL-2008,Ma-arXiv-2008,Si-PRL-2008}
providing $S$ is close to unity. The values of $gS_{\rm eff}$ [0.56
for case (i), 1.08 for case (ii)] obtained from our analysis are
comparable with the measured ordered moment in $\mu_{\rm B}$ of
0.8--0.9 \cite{Huang-arXiv-2008, Su-arXiv-2008}.

Our results show that spin fluctuations in the parent phase of the
pnictides exist over a wide energy range extending up to
$\sim$175\,meV, not much less than that found in the cuprate high
temperature superconductors. Recent neutron scattering studies on
single crystals of SrFe$_{2}$As$_{2}$\cite{Zhao-arXiv-2008} and
CaFe$_{2}$As$_{2}$\cite{McQueeney-arXiv-2008}, though restricted to
energies below 25\,meV, also find steep magnon dispersion relations.
For high $T_{\rm c}$ it is natural to look for a pairing boson with
a large characteristic energy.  The data here show that spin
fluctuations satisfy this requirement. However, if magnetism is to
play a role in the superconducting state then there must also be a
coupling between spin fluctuations and the electrons involved in
pairing. One piece of evidence for this is the recent observation of
a resonant spin excitation in the superconducting state of
Ba$_{0.6}$K$_{0.4}$Fe$_2$As$_2$ \cite{Christianson-arXiv-2008}. More
generally, the itinerant character of magnetism in the pnictides has
been inferred from the nesting of electron and hole Fermi surface
pockets with nesting vector ${\bf Q}_{\rm AF}$
\cite{Dong-EPL-2008,Cvetkovic-arXiv-2008,Kuroki-PRL-2008,Ma-PRB-2008}
and by the observed effect of AFM order on the Fermi surface as
revealed for example by optical spectroscopy
\cite{Dong-EPL-2008,Hu-arXiv-2008}, angle-resolved photoemission
spectroscopy \cite{Yang-arXiv-2008} and quantum oscillations
\cite{Sebastian-arXiv-2008}.

On the other hand, the dynamic magnetic response measured here does
not show any obvious fingerprints of itinerant magnetism, such as
damping due to a Stoner continuum. It will be interesting to follow
the spin excitation spectrum to still higher energies where
itinerant effects generally have the largest influence.\\

We thank I.I. Mazin for helpful comments. This work was supported by
the Engineering and Physical Sciences
Research Council of Great Britain.\\


\end{document}